\documentclass[twocolumn,amsmath,amssymb,prd]{revtex4}

\def\al{\alpha}
\def\be{\beta}
\def\ga{\gamma}
\def\de{\delta}

\def\la{\lambda}

\def\ps{\psi}

\def\La{\Lambda}

\def\fr#1#2{{{#1} \over {#2}}}

\def\frac#1#2{{\textstyle{{#1}\over {#2}}}}

\def\vev#1{\langle {#1}\rangle}

\def\lsim{\mathrel{\rlap{\lower4pt\hbox{\hskip1pt$\sim$}}
    \raise1pt\hbox{$<$}}}
\def\gsim{\mathrel{\rlap{\lower4pt\hbox{\hskip1pt$\sim$}}
    \raise1pt\hbox{$>$}}}
\def\sqr#1#2{{\vcenter{\vbox{\hrule height.#2pt
         \hbox{\vrule width.#2pt height#1pt \kern#1pt
         \vrule width.#2pt}
         \hrule height.#2pt}}}}

\def\pt#1{\phantom{#1}}

\def\vb#1#2{e_{#1}^{{\pt{#1}}#2}}
\def\ivb#1#2{e^{#1}_{{\pt{#1}}#2}}

\newcommand{\beq}{\begin{equation}}
\newcommand{\eeq}{\end{equation}}
\newcommand{\bea}{\begin{eqnarray}}
\newcommand{\eea}{\end{eqnarray}}
\newcommand{\bit}{\begin{itemize}}
\newcommand{\eit}{\end{itemize}}
\newcommand{\rf}[1]{(\ref{#1})}

\begin{document}

\title{Spacetime Symmetry Breaking and Einstein-Maxwell Theory}

\author{Robert Bluhm}

\affiliation{
Physics Department, Colby College,
Waterville, ME 04901 
}

%\date{January 2015}

\begin{abstract}
A vector model with a hybrid form of spacetime symmetry breaking
consisting of explicit diffeomorphism breaking but
spontaneous local Lorentz violation is presented.
The combined effects of these symmetry breakings give rise to
a theory obeying the Einstein-Maxwell equations in a preferred 
spatially homogeneous and isotropic frame, 
with photons emerging as massless Nambu-Goldstone modes.
Interpretations and possible generalizations of this model are discussed, 
and comparisons are made to previous models describing photons as 
Nambu-Goldstone modes.
\end{abstract}

%\pacs{11.30.Cp, 11.10.Ef, 04.40.Nr}

\maketitle

\section{Introduction}

Fundamental spacetime symmetries,
such as diffeomorphism invariance and local Lorentz invariance,
play an essential role in the Standard Model of particle physics
and Einstein's general relativity.
However, a variety of theoretical results stemming from efforts to merge
gravity with quantum physics suggest that these symmetries might
not hold exactly at all energy scales
\cite{cpt,LorentzTests,aknr-tables,ks,akrlmp03,blpr04}.
These include mechanisms found in string theory, 
quantum gravity models, modified gravity theories, alternative
theories describing dark matter and dark energy, and models with
spacetime-varying couplings.

At the level of effective field theory,
described using an action and Lagrangian,
the breaking of these spacetime symmetries involves
a process of either explicit or spontaneous symmetry breaking
\cite{akgrav04,rb15}.
The breaking is explicit if a fixed background field, 
which is nondynamical and does not undergo field variations,
appears directly in the Lagrangian.
Alternatively, if the action remains invariant under a
spacetime symmetry,
but the vacuum solution does not,
then the breaking is spontaneous.
Examples of models with explicit symmetry breaking include 
massive gravity \cite{MGreviews}, 
Chern-Simons gravity \cite{rjsp}, 
and theories with
explicit time-varying couplings \cite{tvar}.
Theories with spontaneous spacetime symmetry breaking involve
a dynamical tensor that acquires a nonzero vacuum value.
Examples include models in which the tensor is 
a vector \cite{ks,akgrav04,rbak,rbffak}, 
a symmetric two-tensor \cite{cmodels}, 
or an antisymmetric two-tensor \cite{phon}.
Models of these types have been used in a wide range of
applications and geometries
\cite{qbak06,akjt,hb14,qbakrx,fins}.

Both of these forms of spacetime symmetry breaking have direct 
physical consequences in field theory and gravity.
For example, with explicit symmetry breaking in a gravitational theory, 
the requirement of general covariance
must be compatible with geometric identities
such as the Bianchi identity as well as the equations of motion
and covariant energy-momentum conservation.
This results in consistency conditions that must hold,
which involve the background field
\cite{akgrav04,rb15}.
Since local symmetries are lost with explicit breaking,
the number of degrees of freedom in a theory can change as well.
On the other hand,
a well known consequence of spontaneous symmetry breaking is that
massless Nambu-Goldstone (NG) modes should appear.
These either propagate as long-range interactions or get reinterpreted through
the Higgs mechanism as degrees of freedom associated with massive gauge fields.

The simplest models involving spontaneous spacetime symmetry breaking
contain a vector field that acquires a vacuum expectation value.
When the NG modes survive as long-range interactions in this type of model,
proposals have been made to interpret them as massless photons.
This idea dates back to models first defined by Bjorken \cite{bjorken63}
using composite fermions and subsequently by Nambu \cite{nambu68}
using a constrained vector field.
These original models were restricted to flat Minkowski spacetime,
and no physical signatures of Lorentz violation were found to emerge.
Instead, it was argued by Nambu that his model was 
equivalent to electromagnetism in a nonlinear gauge.

More general vector models with spontaneous Lorentz violation
incorporate gravity as well as signatures of physical Lorentz violation
\cite{akgrav04,rbak,rbffak,kt02,m05,el05,polyV,bp05,gjw05,clmt06,rbngrpav,cfjn08,cj08,apdt09,ms,bl14,cp15,eu15}.
These types of models are known as bumblebee models.
In these models, a potential $V$ is typically included in the action,
which induces a vacuum expectation value for the vector field.
The potential is formed as a function of a scalar combination $X$ of the 
vector $B_\mu$ and the metric $g_{\mu\nu}$ and possibly other matter fields as well.
The potential has a minimum when $V^\prime = 0$,
where the prime denotes differentiation with respect to $X$.
In the minimum of $V$, the vector has a vacuum value,
denoted as $\vev{B_\mu} = b_\mu$,
It is this background vector that causes local Lorentz
symmetry to be spontaneously broken.

In many bumblebee models,
the potential $V$ is a function of a scalar $X = (B^\mu B_\mu + b^2)$,
where $b$ is a constant with dimensions of mass and $V^\prime = 0$ for $X = 0$.
In this case,
the vacuum vector $b_\mu$ is spontaneously induced as 
a timelike vector obeying $b_\mu b^\mu = - b^2$.
The natural mass scale to arise in an effective theory originating from
mechanisms in string theory or from a quantum theory is the Planck mass.
However,
since Lorentz violation is presumably small,
having escaped detection in high-precision experiments,
additional couplings giving rise to suppressed values for $b$ would  
need to arise as well.

In general,
there are additional degrees of freedom in bumblebee models
compared with Einstein-Maxwell theory
\cite{rbngrpav},
and the possibility that these might arise as ghost modes is an important consideration.
For this reason,
a subset of bumblebee models known as Kosteleck\'y-Samuel (KS) models 
use a Maxwell kinetic term for the vector $B_\mu$
\cite{ks,akgrav04}.
In a linearized limit, 
this eliminates a potential ghost mode as a propagating degree of freedom, 
and in flat spacetime the model appears to be amenable to quantization
\cite{ch14,msa15}.
In the KS model, 
the potential $V$ destroys the local $U(1)$ gauge symmetry that holds for the 
Maxwell kinetic term.
The form of the potential also allows both 
massless NG modes and a massive mode.
The NG modes are excitations about the vacuum solution that remain in the
minimum of the potential obeying $V^\prime = 0$,
while the massive mode is an excitation with $V^\prime \ne 0$.
When a nonzero massive mode is present,
this results in modifications to both the Newtonian and Coulomb static potentials
\cite{rbffak}.
These along with interactions in the matter sector provide physical signals of Lorentz breaking.
However, in the limit where the massive mode becomes extremely large,
the KS model merges with Einstein-Maxwell theory.

In this paper,
a vector model is defined that has a hybrid form of spacetime symmetry breaking.
It uses both explicit diffeomorphism breaking and spontaneous Lorentz breaking.
The idea behind the explicit breaking is to incorporate at the level of effective field theory
the possibility of a spacetime-dependent coupling.
Couplings of this form have been investigated both theoretically and experimentally
\cite{tvar}.
Indeed, one of the original motivations was Dirac's large-number hypothesis
\cite{dirac37},
which suggested that the huge difference in physical scales that are observed in
nature have their origin in the form of a time-dependent coupling.
The coupling considered here is assumed to combine with the mass scale $b$,
giving rise to a model with an explicit time-dependent scalar $b(t)$.
Here, $t$ is presumably time on a cosmological scale,
and the dependence in $b(t)$ could involve a significant suppression factor
compared to the time-independent scale $b$ as the universe has expanded.
With this additional dependence, the potential can have a modified
functional form given as $V(B^\mu B_\mu + b(t)^2)$,
which explicitly breaks time diffeomorphisms.
However, local Lorentz symmetry is still spontaneously broken by this form of $V$
when a vacuum solution obeying $b_\mu b^\mu = - b(t)^2$ appears.

In the next section,
the vector model with a hybrid form of symmetry breaking is presented,
and the consequences of both the explicit time diffeomorphism breaking and
spontaneous Lorentz symmetry breaking are examined.
It is shown that the consistency conditions arising as a result of the explicit
time diffeomorphism breaking impose a constraint on the theory,
which does not allow the massive mode to appear.
The resulting theory therefore only contains massless NG modes as photons
and has equations of motion equivalent to Einstein-Maxwell theory.
Section III offers more detailed interpretations and compares the hybrid form of spacetime
symmetry breaking with spontaneous Lorentz breaking that 
does not involve explicit diffeomorphism breaking.
Possible generalizations of these results are discussed in Sec.\ IV,
and Sec.\ V provides a summary and conclusions.

\section{$b(t)$ Model}

Consider the action
\beq
S = \int d^4 x \sqrt{-g} \, {\cal L} ,
\label{action}
\eeq
where
\bea
{\cal L} &=& \fr 1 {16 \pi G} R + \, {\cal L}_{\rm B} (g_{\mu\nu}, B_\mu)
\nonumber \\
&& 
- V(B_\mu B^\mu + b(t)^2) + {\cal L}_M(g_{\mu\nu}, B_\mu, f^\ps) .
\label{Lagrangian}
\eea
This Lagrangian contains an Einstein-Hilbert term for the metric $g_{\mu\nu}$,
kinetic terms for the vector field in $ {\cal L}_{\rm B}$,
a potential $V(B_\mu B^\mu + b(t)^2)$ that depends on a time-dependent coupling $b(t)$,
and a matter sector $ {\cal L}_{\rm M}$
that couples conventional matter fields denoted generically as $f^\ps$
with the vector and metric.

The kinetic term for the vector is chosen to have a Maxwell form, 
with
\beq
{\cal L}_{\rm B} = - \fr 1 4 B_{\mu\nu} B^{\mu\nu} ,
\label{KS}
\eeq
where $B_{\mu\nu} = D_\mu B_\nu - D_\nu B_\mu$
and $D_\mu$ denotes a spacetime covariant derivative.
It is assumed that there is no torsion, 
so the field strength can
also be written as $B_{\mu\nu} = \partial_\mu B_\nu - \partial_\nu B_\mu$.
The energy-momentum tensors for the vector, potential, and matter sector
are defined, respectively, as
\beq
T_{\rm B}^{\mu\nu} = B^{\mu\al} B^\nu_{\pt{\mu}\al} - \fr 1 4 g^{\mu\nu} B_{\al\be} B^{\al\be} ,
\label{TB}
\eeq
\beq
T_{\rm V}^{\mu\nu} = - V g^{\mu\nu} + 2 V^\prime B^\mu B^\nu ,
\label{TV}
\eeq
\beq
\fr 1 2 \int d^4x \sqrt{-g} \, T_{\rm M}^{\mu\nu} \de g_{\mu\nu}
\equiv  \int d^4x \fr {\de (\sqrt{-g}  {\cal L}_{\rm M})}{\de g_{\mu\nu}} \de g_{\mu\nu} .
\label{TM}
\eeq
Similarly, a current that couples the vector $B_\mu$ with matter can be defined as
\beq
\int d^4x \sqrt{-g} \, J^\mu \, \de B_\mu
\equiv  \int d^4x \sqrt{-g} \fr {\de  {\cal L}_{\rm M}}{\de B_\mu} \, \de B_\mu .
\label{J}
\eeq

In terms of these quantities,
the Einstein equations, the equations of motion for the vector field,
and the matter equations of motion are given, respectively, as
\beq
G^{\mu\nu} = 8 \pi G (T_{\rm B}^{\mu\nu} + T_{\rm V}^{\mu\nu} + T_{\rm M}^{\mu\nu} ) ,
\label{Einsteineq}
\eeq
\beq
D_\mu B^{\mu\nu} = 2 V^\prime B^\nu - J^\nu ,
\label{Beq}
\eeq
\beq
\int d^4x \sqrt{-g} \, \fr {\de  {\cal L}_{\rm M}}{\de f^\ps} \, \de f^\ps = 0 .
\label{Meq}
\eeq
Taking covariant divergences
of the first two of these and using the contracted Bianchi identity,
$D_\mu G^{\mu\nu} = 0$, and the equation $D_\mu D_\nu B^{\mu\nu} = 0 $
gives the conditions,
\beq
D_\mu (T_{\rm B}^{\mu\nu} + T_{\rm V}^{\mu\nu} + T_{\rm M}^{\mu\nu} ) = 0 ,
\label{Tconsv}
\eeq
\beq
D_\mu (2 V^\prime B^\mu - J^\mu ) = 0 .
\label{Jconsv}
\eeq
Notice that $V^\prime$ in these equations denotes the appearance of a massive mode.
It acts as both a source of current and energy-momentum density.

The action $S$ is not invariant under time diffeomorphisms.
The change in the total action under these transformations
is obtained by taking field variations of $S$ with respect to the
dynamical fields where the field fluctuations are given by Lie derivatives.
However, the background $b(t)$ is nondynamical and is fixed
under these transformations.
As a result, the Lagrangian does not transform as a scalar
under diffeomorphisms, 
and the action is not invariant.

Despite the explicit breaking of time diffeomorphisms, 
the action is invariant under local Lorentz transformations. 
In some respects this is surprising, 
since a fixed nonconstant background $b(t)$
has field gradients associated with it.
These give preferred directions in local frames,
which as fixed backgrounds do break local Lorentz invariance.
Nonetheless, the action does not depend on the gradient of $b(t)$,
and therefore there is no explicit breaking of local Lorentz symmetry
in the resulting dynamics as described by $S$.

However, there is spontaneous breaking of local Lorentz invariance,
due to the form of the potential $V$,
which causes nonzero vacuum values to appear.
Assuming vanishing kinetic terms,
the vacuum solution consists of a vector $b_\mu$
obeying $b_\mu \vev{g^{\mu\nu}} b_\nu = - b(t)^2$
and a vacuum solution for the metric denoted as $\vev{g_{\mu\nu}}$.
It is assumed that the conventional matter fields have vanishing vacuum values $\vev{f^\ps} = 0$.
These vacuum values involving a timelike and time-dependent vector select a preferred
frame in which the vector takes the form $b_\mu = (b(t),0,0,0)$.
The metric can be written generically as $\vev{g_{\mu\nu}} = {\rm Diag} (-1, a(t), a(t), a(t))$,
with $a(t)$ a scale parameter,
which describes a spatially homogeneous and isotropic vacuum.

When diffeomorphism invariance is explicitly broken by a background field there are potential
inconsistencies that must be overcome for solutions to exist.
The extent to which a theory is constrained by
these consistency conditions depends on the form of the background tensor
\cite{rb15}.
For scalar fields, where the Lie derivative is proportional to the
transformation vector $\xi^\mu$,
e.g., ${\cal L}_{\xi} b(t) = \xi^\mu \partial_\mu b(t)$,
the conditions are most severe.  
In some cases solutions can be ruled out,
while in others solutions can exist only if certain constraints hold.
An example of this type is Chern-Simons gravity,
which has a nondynamical scalar background that explicitly breaks diffeomorphisms.
Solutions to this theory can only exist if the spacetime has
a vanishing Pontryagin density, $^*R R = 0$
\cite{rjsp}.
However, for theories with background vectors or tensors
the Lie derivatives also contain terms that involve derivatives of $\xi^\mu$.
In variations of the action,
this allows integrations by parts to be performed,
which leads to more options for evading the potential inconsistency.
An example along these lines is massive gravity,
which contains a background field that is a symmetric two-tensor.

The vector theory presented here has a background scalar $b(t)$
that appears as part of the potential $V(B_\mu B^\mu + b(t)^2)$.
The requirement of general covariance can be used to determine the
consistency conditions that arise in this case.
Although the $b(t)$ model explicitly breaks time diffeomorphisms,
it must still be generally covariant under coordinate transformations
to maintain observer independence.
This includes general coordinate transformations defined as
$x^\mu \rightarrow x^\prime (x) = x^\mu - \xi^\mu$,
which have the same mathematical form as a diffeomorphism transformation.
However, under observer general coordinate transformations 
the background $b(t)$ transforms, 
and the Lagrangian is therefore a scalar under these transformations.

Performing the field variations corresponding to these observer transformations
on $S$ gives the off-shell result:
\bea
&& \int d^4 x \,  \left[
 \fr {\de (\sqrt{-g} ({\cal L_{\rm B}}  + {\cal L_{\rm M}} - V)} {\de g_{\mu\nu}} {\cal L}_\xi g_{\mu\nu}
 \right.
 \quad\quad\quad \quad\quad\quad
 \nonumber \\
 && \quad
 + \sqrt{-g} \fr {\de ({\cal L}_{\rm B} + {\cal L_{\rm M}} - V)} {\de B_\mu} {\cal L}_\xi B_\mu 
+ \sqrt{-g} \fr {\de {\cal L}_{\rm M}} {\de f^\ps} {\cal L}_\xi f^\ps 
 \nonumber \\
 && \quad\quad\quad\quad\quad\quad\quad\quad
 \left.
 - \sqrt{-g} \fr {\de V} {\de b(t)} {\cal L}_\xi b(t) 
 \right]
 = 0 .
 \label{covariancecond}
\eea
In this expression, the variations of the Einstein-Hilbert term drop out as
a result of integrating by parts and using the contracted Bianchi identities.
Since the fields $B_\mu$ and $f^\ps$ are dynamical
their variations in the action vanish on-shell.
The remaining variations with respect to the metric $g_{\mu\nu}$
define the energy-momentum tensors.
Using integrations by parts on these terms gives the result,
\bea
\int d^4 x \,  \sqrt{-g}  \, \xi_\nu \left[ 
- D_\mu (T_{\rm B}^{\mu\nu} + T_{\rm V}^{\mu\nu} + T_{\rm M}^{\mu\nu} )
\right.
\nonumber \\
\quad\quad
\left.
 - \fr {\de V} {\de b(t)} D^\nu b(t) 
 \right]
 = 0 .
 \label{covariancecond2}
\eea
This result must hold for all $\xi_\nu$ with appropriate boundary conditions.
For the case where the potential has the form $V(B_\mu B^\mu + b(t)^2)$,
the resulting consistency condition is
\beq
D_\mu (T_{\rm B}^{\mu\nu} + T_{\rm V}^{\mu\nu} + T_{\rm M}^{\mu\nu} ) = 
- 2 V^\prime b(t) D^\nu b(t) .
\label{DTtotal}
\eeq

If the right-hand side in this expression is nonzero,
this result is clearly in conflict with the condition in Eq.\ \rf{Tconsv},
which followed from the Einstein equations and the contracted Bianchi identity.
Thus, the right-hand side must vanish for solutions to exist, giving
\beq
V^\prime \, b(t) \, \partial_0 b(t) = 0 .
\label{cond}
\eeq
Since $\partial_0 b(t) \ne 0$ by construction,
the result is that the massive mode $V^\prime$ must vanish for solutions to exist.
With $V^\prime = 0$, the only excitations that can exist for the
vector field are the massless NG modes.

This same result follows from the equations of motion as well.
Using the definitions of $T_{\rm B}^{\mu\nu}$ and $T_{\rm V}^{\mu\nu}$,
their divergences can be worked out and combined with the
equations of motion.
The results are
\beq
D_\mu T_{\rm B}^{\mu\nu} = -2 V^\prime B_\mu B^{\mu\nu} + J_\mu B^{\mu\nu} ,
\label{DTB}
\eeq
\bea
D_\mu T_{\rm V}^{\mu\nu} &=& 2 V^\prime B_\mu B^{\mu\nu} + (D_\mu J^\mu) B^\nu 
\nonumber
\\
&& \quad\quad - 2 V^\prime b(t) D^\nu b(t) .
\label{DTV}
\eea
For the matter sector, a specific form for the fields $f^\mu$ is required.
However, a covariance argument can be used for the term ${\cal L}_{\rm M}$, 
with the requirement that it by itself must be a scalar.
The result is
\beq
D_\mu T_{\rm M}^{\mu\nu} = - (D_\mu J^\mu) B^\nu - J_\mu B^{\mu\nu} .
\label{DTM}
\eeq
Adding these three expressions gives \rf{DTtotal}.

Notice that in the individual expressions for the different contributions
to the energy-momentum density,
when a massive mode is present with $V^\prime \ne 0$,
then matter charge current density is not conserved
and exchanges of energy depending on the massive mode and the
matter charge nonconservation can occur between the different sectors.
It is these types of transfers that can destabilize the theory if the massive
mode is not constrained.

However, the condition $V^\prime = 0$ that must hold on shell
alters the equations of motion and conservation conditions.
With no massive mode,
the equation of motion for the vector field in \rf{Beq} 
reduces to the usual Maxwell equations.
Taking the divergence then gives
\beq
D_\mu J^\mu = 0 .
\label{DJ0}
\eeq
With $V^\prime = 0$,
the energy-momentum $T_{\rm B}^{\mu\nu}$ becomes
equivalent to the energy-momentum in electromagnetism,
and $T_{\rm V}^{\mu\nu}$ reduces to a contribution from a 
cosmological constant with $V$ equal to a constant.
Notice that with $V^\prime = 0$,
the only exchanges of energy between the vector field
and the matter sector have the usual form as a Lorentz force 
$\pm J_\mu B^{\mu\nu}$.

In the $b(t)$ model,
the only vector excitations are massless NG modes,
which are solutions of the usual Einstein-Maxwell equations,
but with a fixed gauge determined by the condition $V^\prime = 0$.
If $V^\prime = 0$ is satisfied by 
$B_\mu B^\mu = - b(t)^2$,
then the NG modes are excitations that preserve this condition.
If the theory is linearized, using $B_\mu \simeq b_\mu + {\cal E}_\mu$,
then to leading order the condition is satisfied by
excitations obeying an axial gauge condition, $b^\mu {\cal E}_\mu = 0$.
These excitations can be shown to consist of two transverse massless modes
and one auxiliary mode,
which is the same as for a massless photon.

Since $B_\mu$ reduces to the background $b_\mu$ in the absence of photons, 
interactions with matter currents of the form
$b_\mu J^\mu$ might be expected to cause Lorentz-violating signals,
which would be of a form as described by the
Standard-Model Extension (SME)
\cite{sme,rbsme}.
This would then lead to experimental bounds being placed on $b_\mu$.
However,
interactions of this form with a fermion matter field
are known to be unobservable in the SME.
This is because the coefficients $b_\mu$ can be absorbed
by a field redefinition that shifts the phase of the fermion field.
In the absence of the massive mode,
there are also no modifications of the Newtonian or Coulomb
potentials as there are in the KS model.
The end result appears to be that the model with a background
time-varying coupling $b(t)$ gives solutions that are equivalent to Einstein-Maxwell
theory in a nonlinear gauge.

Notice that this type of theory has no analogue in flat spacetime.
While a flat background is a valid vacuum solution,
the spacetime itself must remain dynamical,
allowing gravitational excitations to occur.  
This is because the condition $V^\prime = 0$ is imposed as a result
of combining the contracted Bianchi identity with the Einstein equations.
It therefore hinges on the dynamics and geometrical conditions that apply in gravity.
However, in the absence of gravity, no such condition emerges and $V^\prime$ is
not required to vanish.
As a result, the theory in flat spacetime violates energy conservation
due to the breaking of time translation invariance.

\section{Interpretations and Comparisons}

The results obtained for the $b(t)$ model raise a number of interpretational issues,
which are discussed in this section.
For many of these, 
it is useful to make comparisons with the KS model and
also with Einstein-Maxwell theory.  

The KS model has a constant value of $b$,
and there is no explicit symmetry breaking.
Instead,
time diffeomorphisms and local Lorentz symmetry are spontaneously 
broken by a constant background,
which can be chosen in a preferred frame as $b_\mu = (b,0,0,0)$.
There is no condition that the massive mode $V^\prime$ must vanish.
The NG modes appear as photons;
however,
there are also signatures of Lorentz violation
due to the presence of the massive mode.

\subsection{Degrees of Freedom}

As a first comparison,
the number of degrees of freedom can be examined
for both the KS and $b(t)$ models.
Since the KS model is diffeomorphism invariant,
there are four gauge degrees of freedom associated with this symmetry.
However, due to the breaking of local $U(1)$ gauge invariance
there is one additional degree of freedom compared to Einstein-Maxwell theory,
which is the massive mode $V^\prime$.
This is the case in the $b(t)$ model as well.
However,
due to the breaking of time diffeomorphisms in the $b(t)$ model,
there is a second additional degree of freedom in the form of the
metric component that can no longer be gauged away.

The equations of motion associated with these extra degrees of freedom
can be investigated at the linearized level using field redefinitions.
For the broken $U(1)$,
this is achieved by writing $B_\mu \simeq A_\mu - \partial_\mu \La$,
where $A_\mu$ is a gauge-fixed vector,
for example satisfying an axial gauge condition,
while $\La$ is the extra degree of freedom associated with the broken local $U(1)$ symmetry.
Substituting this in the Lagrangian
and varying with respect to $\La$ gives the equation
$D_\mu (2 V^\prime B^\mu - J^\mu) = 0$.
This holds as a dynamical equation of motion in both
the KS and $b(t)$ models.
The additional equation in the $b(t)$ model can
be obtained in a similar way at the linearized level
by writing $g_{\mu\nu} \simeq \tilde g_{\mu\nu} + \de_\nu^0 D_\mu \xi_0 + \de_\mu^0 D_\nu \xi_0$,
where $ \tilde g_{\mu\nu}$ is a gauge-fixed form of the metric and
$\xi_0$ is the extra degree of freedom associated with the broken time diffeomorphism.
In this case, varying the effective action with respect to $\xi_0$ gives
$D_\mu (T_{\rm B}^{\mu 0} + T_{\rm V}^{\mu 0} + T_{\rm M}^{\mu 0} ) = - 2 V^\prime b(t) D^0 b(t)$.
Note that when the equation for $\xi_0$ is combined with the Einstein
equations and the contracted Bianchi identity the end result is that $V^\prime = 0$.
This additional condition does not arise in the KS model because it has only the
one additional equation of motion associated with the broken $U(1)$ symmetry.

\subsection{Current Conservation}

Another feature of the $b(t)$ model compared to the KS model is
that the charge current $J^\mu$ must be covariantly conserved in the $b(t)$ model, 
while it need not be in the KS model if a massive mode is present.
The existence of a conserved current in the $b(t)$ model suggests
that there should be a $U(1)$ symmetry.
However, this symmetry is explicitly broken in the $b(t)$ model.
Thus, the question of why covariant current conservation holds in
the $b(t)$ model needs to be addressed.

To begin, note that while local $U(1)$ symmetry and time diffeomorphisms are
broken by the potential term in the $b(t)$ model,
there is an unbroken diagonal subgroup.
This can be used to show that if a generic external current
$J^\mu$ couples to $B_\mu$,
with ${\cal L}_{\rm M} = B_\mu J^\mu$,
then $J^\mu$ must be conserved in order for
the subgroup symmetry to hold.
To see this,
transform the action by
both a broken infinitesimal time diffeomorphism with vector $\xi^0$
and by a broken local $U(1)$ transformation,
$B_\mu \rightarrow B_\mu - \partial_\mu \La$.
It is assumed that the only breaking of these symmetries in in the potential term.
Therefore,
to leading order in the infinitesimal parameters, the result is
\beq
\de S = \int d^4 x \sqrt{-g} 
V^\prime \left( - 2 B^\mu \partial_\mu \La + 2 b(t) \xi^0 \partial_0 b(t) \right) .
\label{deS}
\eeq
This is an off-shell result showing that a local subgroup symmetry exists if $\La$ and $\xi^0$
are chosen at every point so that the term in parentheses vanishes.
The entire action is then invariant under the subgroup transformation,
which implies that the matter term ${\cal L}_{\rm M}$ 
by itself is invariant under just the $U(1)$ transformation, 
since $b(t)$ does not enter ${\cal L}_{\rm M}$.
Performing the subgroup transformation,
which leaves the total action invariant, 
therefore gives
${\cal L}_{\rm M} \rightarrow {\cal L}_{\rm M} - (\partial_\mu \La) J^\mu$ in the matter term.
The vanishing of this extra term in the action requires that $D_\mu J^\mu = 0$ must hold.

An example of how current conservation arises in interactions with
a specific type of matter field can be considered as well.
Consider the case of a fermion field that is minimally coupled to $B_\mu$.
To describe gravity with a fermion field,
a vierbein formalism is used,
where the metric is replaced by a vierbein $\vb \mu a$.
Here, the index $a$ gives components defined with respect to a
local Lorentz frame,
and the Dirac form of the action can then be used.
In the absence of torsion, as assumed here,
the vierbein formalism does not
play a significant role in the conservation law 
that arises for the current.
The main effect is that the Dirac matrix $\ga^a$ in a local Lorentz frame
becomes the composite $\ivb \mu a \ga^a$ in a curved spacetime.
The minimally coupled matter term for a fermion $\ps$ is then given as
\beq
{\cal L}_{\rm M} = \bar \ps \left(i \ivb \mu a \ga^a (D_\mu + i q B_\mu) - m \right) \ps  .
\label{Dirac}
\eeq
In this case, the term ${\cal L}_{\rm M}$ by itself is invariant under local 
$U(1)$ tranformations,
since the current is carried by the dynamical fermion fields.
However, the theory also has a global $U(1)$ symmetry,
$\ps \rightarrow e^{iq\La} \ps$, where $\La$ is a constant
and all other fields are left unchanged.
Applying the Noether theorem to this global symmetry
gives a covariantly conserved current $J^\mu = \bar \ps \ivb \mu a \ga^a \ps$,
which obeys $D_\mu J^\mu = 0$ on shell.
Note that this current does not depend on the potential $V$,
which is independent of $\ps$.
As a result, the current $J^\mu$ carried by the fermion fields
is conserved in both the KS and $b(t)$ models.
Note as well that the condition $D_\mu J^\mu = 0$,
causes the charge current associated with the massive mode
to be separately conserved, 
i.e., $D_\mu (2 V^\prime B^\mu) = 0$ must hold on shell in either model.
However, this condition does not require that $V^\prime = 0$,
and indeed in the KS model the massive mode $V^\prime$ does not need to vanish.
Similarly, in the $b(t)$ model,
current conservation for $J^\mu$ is not
sufficient for requiring that the massive mode must vanish.

Since the unbroken diagonal subgroup is a 
local symmetry of the action,
there should be a Noether identity associated with it that holds on shell.
However, this identity would not give any additional conditions 
that do not already follow from the Bianchi identity,
the equations of motion, and the identity $D_\mu D_\nu B^{\mu\nu} = 0 $.
It is the consistency of these conditions that requires that $V^\prime = 0$ 
must hold in the $b(t)$ model.

\subsection{Einstein-Maxwell with Gauge-Fixing Term}

The question of whether the potential term $V(B^\mu B_\mu + b(t)^2)$ 
in the $b(t)$ model can be considered as equivalent to 
a gauge-fixing term in Einstein-Maxwell theory can be addressed as well.
In an interpretation along these lines,
the initial Lagrangian is taken as the usual Einstein-Maxwell Lagrangian,
and the potential is treated as an added gauge-fixing term that is used to fix a particular gauge choice.
The most straightforward way to implement a fixed gauge is by using a Lagrange-multiplier potential.
As an illustrative comparison,
first consider the KS model with a Lagrange-multiplier potential,
which is given as
\beq
V = \la (B_\mu B^\mu + b^2) ,
\label{LagrangeV}
\eeq
where $\la$ is the Lagrange-multiplier field.
Variation of the action with respect to $\la$ in this case gives the equation $B^\mu B_\mu = -b^2$,
which can be interpreted as a nonlinear gauge-fixing condition for the local $U(1)$ symmetry.
The only excitations permitted in $B_\mu$ are the NG modes.
However, the massive mode $V^\prime = \la$ still appears in the remaining equations of motion,
and equivalence with Einstein-Maxwell theory is only achieved when $\la = 0$.
Thus, an essential part of the gauge-fixing procedure is to get rid of the Lagrange multiplier
by setting $\la = 0$ by hand.
There is no dynamical condition in the KS model that requires that $\la$ must vanish.
It is for this reason that the KS model is not equivalent to Einstein-Maxwell theory
with a gauge-fixing term.

In contrast, the $b(t)$ model with a Lagrange-multiplier potential
$V= \la (B^\mu B_\mu + b(t)^2)$ has $B^\mu B_\mu = -b(t)^2$
arising as the equation of motion for $\la$.
However, in this case, the requirement of general covariance leading to Eq.\ \rf{DTtotal} still applies,
and this combined with the contracted Bianchi identity and the
equations of motion gives \rf{cond},
but with $V^\prime = \la$.
Since $\partial_0 b(t) \ne 0$ by construction,
the consistency of the theory requires that $\la = 0$ must always hold on shell.
Thus, 
in the case of the $b(t)$ model with a Lagrange-multiplier potential,
the condition $B^\mu B_\mu = -b(t)^2$ is automatically imposed by the consistency conditions 
that follow as a result of explicit diffeomorphism breaking.
It does not involve an extra procedure that must be implemented by hand.
Moreover, for any potential of the form $V(B^\mu B_\mu + b(t)^2)$,
even without a Lagrange-multiplier field,
the consistency of the theory requires that $V^\prime = 0$ must hold.
There are therefore an infinite number of possible terms that all give similar results
equivalent to Einstein-Maxwell theory with the same fixed nonlinear gauge condition.
It is important to note as well that the potentials $V(B^\mu B_\mu + b(t)^2)$ 
do not fully fix the broken gauge symmetries,
since a local residual subgroup symmetry still exists.  

Ultimately, if the $b(t)$ model is fully equivalent to Einstein-Maxwell theory 
then it makes sense that it would be open to more than one form of interpretation.
If the model is viewed as an Einstein-Maxwell theory with a gauge-fixing term,
it is fair to say that it involves an unusual choice of gauge,
which is both nonlinear and spacetime dependent.
Nonetheless, with this interpretation, 
photons can be considered as massless gauge fields.
On the other hand, if the $b(t)$ model is interpreted as a theory with a
hybrid form of spacetime symmetry breaking,
then photons emerge in this case as massless NG modes
associated with spontaneous Lorentz violation.
In this interpretation, the result that only the NG modes can appear
(with no massive mode) is a consequence of the conditions that must hold when 
diffeomorphisms are explicitly broken in a gravitational theory.  

\subsection{Symmetry Breaking}

Lastly, there are issues concerning how to interpret the symmetry breaking in the $b(t)$ model,
which merit discussion as well.
The vacuum solution involves a vector $b_\mu$ that satisfies the
condition $b^\mu b_\mu = - b(t)^2$.
In this relation, $b(t)$ is a fixed background that explicitly breaks time diffeomorphisms.
On the other hand,
$b_\mu$ is a vacuum expectation value that
appears as a result of spontaneous symmetry breaking.
It spontaneously breaks local Lorentz boosts as well as time diffeomorphisms.  
Thus, there appears to be a double breaking of time diffeomorphisms,
once explicitly by $b(t)$ and then again spontaneously by $b_\mu$.

However, notice that the consistency condition stemming from
explicit diffeomorphism breaking requires that the last term in 
Eq.\ \rf{covariancecond} must vanish.
It is this condition that gives $V^\prime = 0$ as an on-shell
condition in the $b(t)$ model.
The vanishing of this term,
involving variation of the action with respect to $b(t)$, 
is the same result that would hold if $b(t)$ were in fact a dynamical field.
This feature of explicit diffeomorphism breaking by a fixed background scalar
was also observed in Chern-Simons gravity
\cite{rjsp}.
For consistency to hold,
the background $b(t)$ must act effectively like a dynamical solution.
However, since $b(t)$ is fixed,
this is actually a condition that is imposed on the other fields,
i.e., $B_\mu$ and $g_{\mu\nu}$.
It is these fields that must have solutions that permit $b(t)$ to blend in as a dynamical solution.
It is for this reason that a constraint gets put on $B_\mu$ and $g_{\mu\nu}$.

With $b(t)$ effectively mimicking a dynamical field,
the explicit breaking and spontaneous breaking
of time diffeomorphisms are compatible.
In particular, when a vacuum solution forms,
the explicit-breaking background $b(t)$ coexists with 
$b_\mu$ and $\vev{g_{\mu\nu}}$
as if it too were a dynamical vacuum solution.

\section{Generalizations}

There are a number of generalizations of the $b(t)$ model that can be considered.
The key feature in these is the inclusion of fixed backgrounds that impose conditions
as a result of explicit diffeomorphism breaking.

One generalization would be to use a theory containing higher-rank tensors.
In this case, a potential $V$ would be constructed out of scalars formed
using these tensors and the metric.
Possible examples include using a symmetric two-tensor or an
antisymmetric two-tensor,
both of which are used in models with spontaneous Lorentz breaking
\cite{cmodels,phon}.
Generalizing these to allow potentials that have background fields that
explicitly break diffeomorphisms might give new theories with hybrid
forms of spacetime symmetry breaking.
Combining broken diffeomorphisms with broken gauge symmetry groups
may result in theories with unbroken subgroups.
With higher-rank tensors,
it becomes possible to explicitly break more than one diffeomorphism,
which can give consistency conditions that require additional constraints.

Further modifications to a vector theory with explicit breaking can be considered as well.
For example, vector fields with a nonabelian gauge group might be used
\cite{cj08,eu15}.
Alternatively, different forms of kinetic terms ${\cal L}_{\rm B}$ besides 
the Maxwell form could be included.
Models with generalized kinetic terms are typically investigated as vector-tensor theories
\cite{wn73}
of gravity as opposed to modified theories of electromagnetism,
and for this reason they typically do not include direct matter couplings.
The symmetry breaking in this case might lead to modified forms of
propagation of gravitational interactions,
making the interpretation in terms of spontaneous Lorentz breaking
with generation of a vacuum solution and NG modes less relevant.  
Such models with explicit breaking can also result in modified initial value constraints
\cite{pond}.
Another modification would be to include torsion
with a dynamical spin connection
\cite{akgrav04}.
This allows the possibility of a Higgs mechanism
in a Riemann-Cartan geometry
\cite{rbak,rbffak}, 
where a propagating spin connection can acquire a mass.   
With additional couplings that explicitly break diffeomorphism invariance,
other forms of Higgs approaches in gravity \cite{Higgs}
might emerge.
Nonmiminal gravitational couplings of the vector $B_\mu$ with the
curvature tensor can be considered as well
\cite{akjt}.
These types of couplings are known to give rise to physical signals of 
spontaneous Lorentz violation.
In all of these modifications,
the possible generation of ghost modes becomes a serious problem
that would need to be overcome to obtain viable models.

Since the $b(t)$ model is based on the idea that a time-varying
coupling can arise at the level of effective field theory from 
unknown mechanisms occurring in the context of a more fundamental theory,
it is possible that more than one such coupling might arise.
For example, while an explicit time-varying cosmological
constant $\La (t)$ by itself is inconsistent with the Einstein
equations, the Bianchi identity, and matter energy-momentum conservation,
this would no longer be the case when other time-varying
couplings are included.
In the $b(t)$ model,
adding a time-varying cosmological constant would lift the 
requirement that the massive mode must vanish.
Instead, the time variations of $b(t)$ and $\La(t)$ would
be linked by the consistency conditions associated 
with explicit diffeomorphism breaking.
However, extended models of this form lack
a clear determination of the functional
time dependence that appears in these couplings.
Such an issue is less of a concern in the $b(t)$ model
that does not include additional time-dependent couplings, 
since ultimately the theory is found to be equivalent to Einstein-Maxwell theory.

\section{Summary and Conclusions}

This paper considers the idea that an effective field theory
arising from a more fundamental theory at the Planck scale,
such as string theory or a quantum theory of gravity, 
might incorporate both spontaneous Lorentz violation 
and the formation of a time-varying coupling.
Such a theory would have a hybrid form of spacetime
symmetry breaking consisting of both explicit diffeomorphism
breaking and spontaneous Lorentz breaking.
Each of these types of symmetry breaking has 
physical consequences,
which are explored for the case of a gravitational theory 
with a vector field.

The model with a hybrid form of symmetry breaking 
considered in this paper replaces the constant
$b$ in the KS model with a time-varying coupling $b(t)$.
The resulting potential then has the form $V(B^\mu B_\mu + b(t)^2)$,
where it is assumed that the minimum of the potential
occurs when $B^\mu B_\mu = - b(t)^2$.
The appearance of $b(t)$ in the effective Lagrangian explicitly
breaks time diffeomorphisms,
but still allows spontaneous breaking of local Lorentz symmetry.
The potential also explicitly breaks local $U(1)$ symmetry;
however, an unbroken diagonal symmetry remains,
which manifests itself in the matter sector as a local $U(1)$ transformation.

The explicit breaking of time diffeomorphism invariance
results in consistency conditions that must hold on shell
\cite{akgrav04,rb15}.
These conditions stem from the combination of general coordinate invariance,
the dynamical equations of motion, and the Bianchi identity.
In the $b(t)$ model, they require that the only allowed solutions 
are ones that keep the potential at its minimum with $V^\prime = 0$.
This forbids the appearance of a massive mode and only allows
the NG modes as possible excitations of the vector around its vacuum solution.
With $V^\prime = 0$, the equations of motion are equivalent to
those in Einstein-Maxwell theory,
and both the total energy-momentum tensor and the charge current
are covariantly conserved on shell.
Thus, the NG modes appear as photons.

The vacuum solution that results has a preferred frame in
which the background $b_\mu$ is purely timelike,
and the vacuum solutions for the metric describe a 
spatially homogeneous and isotropic spacetime.
With just one vector background $b_\mu$,
there are no conventional interactions with matter fields
that cannot be eliminated using field redefinitions.
Thus, the $b(t)$ model does not have physical signatures of Lorentz violation.
Instead, the hybrid form of spacetime symmetry breaking
can be viewed as an alternative explanation for the emergence of massless photons
in a classical gravitational field theory besides the usual one based on gauge invariance.

%\section*{Acknowledgments}

\end{document}